\documentclass[a4]{article}
\usepackage{amsmath}
\allowdisplaybreaks

\usepackage{thophys}

\begin{document}
\newcommand{\xib}{\overline{\xi}}
\renewcommand{\thefootnote}{\fnsymbol{footnote}}
\thispagestyle{empty}
\begin{titlepage}
\begin{flushright}
hep-ph/9809226 \\
TTP 98--30 \\
\today
\end{flushright}

\vspace{0.3cm}
\boldmath
\begin{center}
\Large\bf  Reparametrization Invariance of Heavy Quark Effective Theory at
           $\mathcal O(1/m_Q^3)$
\end{center}
\unboldmath
\vspace{0.8cm}

\begin{center}
{\large Christopher Balzereit\footnote[5]{email:
    chb@particle.physik.uni-karlsruhe.de}}\\
{\sl Institut f\"{u}r Theoretische Teilchenphysik,
     Universit\"at Karlsruhe,\\ D -- 76128 Karlsruhe, Germany} 
\end{center}

\vspace{\fill}

\begin{abstract}
\noindent
We extend the investigation of the reparametrization invariance
of Heavy Quark Effective Theory to $\mathcal O(1/m_Q^3)$.
We show that in the presence of radiative corrections 
reparametrization invariance
can only be maintained if the the reparametrization transformation 
of the fields is renormalized properly.
\end{abstract}
\end{titlepage}

\renewcommand{\thefootnote}{\arabic{footnote}}
\newpage

\section{Introduction}
\label{sec:introduction}
Heavy Quark Effective Field Theory
(HQET)~\cite{HQET_classic} has become a well
established theoretical tool for the description of
hadrons containing one heavy
quark~\cite{HQET_review}.  This derives from the fact that it is a
systematic expansion in inverse powers of the heavy quark mass $m_{Q}$ with
well defined and calculable coefficients.  Furthermore, its
realization of the spin and flavor symmetry of the low energy theory
is a phenomenologically powerful tool. 
The $1/m_Q$ expansion has already been applied sucessfully to 
problems such as the determination of $V_{cb}$. 

Besides its phenomenological application HQET posseses 
interesting theoretical features which already show up if one studies the HQET 
lagrangian itself.  
On one hand it is relevant for power corrections
to the $1/m_{Q}$ expansion of hadron masses and on the other hand allows for 
a study of reparametrization invariance \cite{repar}. 

This new symmetry of HQET is associated with the fact, 
that QCD Greensfunctions 
only depend on the full heavy quark momentum 
$p = m_{Q}v + k$ independent
from its decomposition  in on shell momentum $mv$ and residual momentum
$k$ in terms of which HQET is defined. 
This means, that an infinitesimal change of the velocity $v \to v + \delta v$ 
can be compensated by an appropriate change of the residual momentum
$k \to k - m_{Q} \delta v$ corresponding to 
a transformation $h_{v} \to h_{v} + \delta h_{v}$ of the 
heavy quark field. 
As long as $m\delta v = \mathcal O(\Lambda_{QCD})$
this change of the velocity does not violate the condition $k = \mathcal
O(\Lambda_{QCD}))$ on which the construction of HQET relies.

There is no unique way to define HQET since a
redefinition 
of the heavy quark field 
leads to an equivalent formulation of
HQET due to the equivalence theorem.
The equivalent formulations of HQET differ from each other by 
operators vanishing by the equation of motion (EOM) of the 
heavy quark. For example
in the Foldy--Wouthuysen formulation 
which is used in NRQCD applications
these 
operators serve to remove all covariant time derivatives
from the lagrangian. However, if these unphysical 
operators are removed consistently, one ends with
a unique minimal lagrangian consisting only of physical operators.
By definition a physical operator contains 
no $(ivD)h_v$-- or $\bar h_v (ivD)$--term.
In general the reparametrization transformation of the
heavy quark field depends on what specific 
formulation of HQET has been choosen.

Given the transformation $\delta h_{v}$ associated with a certain formulation
of HQET at tree level it is easy to verify that the corresponding 
lowest order lagrangian 
is reparametrization invariant. 
However the question arises if the tree level transformation suffices
if radiative corrections are included in the
lagrangian. From usual symmetries, e.g. gauge symmetries, 
it is well known that in each order of perturbation theory the
BRS--transformations and the Slavnov identities 
have to be renormalized. 
Accidentally this seems not to be the case for reparametrization invariance
at least if power corrections higher than $\mathcal O(1/m_{Q}^{2})$
are excluded. Several authors \cite{blok,grozin,finke} 
have verified explicitly
that to this order reparametrization invariance holds
if the tree level transformation is used even in the
presence of radiative corrections.
In \cite{manohar} reparametrization invariance 
of the NRQCD lagrangian 
at $\mathcal O(1/m_{Q}^{3})$ has been studied.
However, a direct comparison with our results is not possible
since we are working in different HQET--formulations.

Recently all coefficients of the effective lagrangian at $\mathcal
O(1/m_{Q}^{3 })$ including EOM--operators 
have become available \cite{balzereit}. 
It is the subject of this 
paper to extend the analysis of reparametrization invariance to
this order.  It will be shown, that the above naive understanding 
of reparametrization invariance fails and how the emerging inconsistencies
can be cured. 

This paper is organized as follows: In section \ref{sec:lag} we 
present the effective lagrangian in the formulation
introduced by Mannel, Roberts and Ryczak (MRR) \cite{MRR} 
and the field redefinition which
removes the operators vanishing by the equation of motion.
In section \ref{sec:RI} we introduce 
the concept of reparametrization invariance and
derive a relation between reparametrization transformations
associated with different formulations of HQET. 
Sections \ref{sec:RIMRR} and \ref{sec:RIphys} apply
this formalism to the MRR lagrangian and the minimal lagrangian.
Finally we present our conclusions in section \ref{sec:conc}.
\section{MRR--HQET up to $\mathcal O(1/m_Q^3)$}
\label{sec:lag}
HQET can be derived from the QCD lagrangian
\begin{equation}
\mathcal L = \bar Q (i\fmslash{D} - m_Q) Q + \ldots
\end{equation}
by reparametrization 
of the heavy quark field $Q$ in terms of its particle and antiparticle
 components: 
\begin{equation}\label{eq:match}
Q = e^{-im_{Q}vx}(h_v + H_v), \quad \fmslash{v}h_v = h_v, 
\quad \fmslash{v}H_v = -H_v 
\end{equation}
In order to formulate an effective theory for the particle degree of freedom
$h_{v }$ the heavy degree 
of freedom $H_{v}$ has to be integrated out. 
This yields the nonlocal lagrangian
\begin{equation}\label{eq:lagMRRnl}
\mathcal L = \bar h_v  (ivD) h_v
 + \bar h_v i\fmslash{D}^{\bot} \frac{1}{ivD + 2m_Q} i\fmslash{D}^{\bot} h_v\,.
\end{equation}
Accordingly the matching condition (\ref{eq:match}) becomes nonlocal:
\begin{equation}\label{eq:matchnl}
Q = e^{-im_{Q}vx}(1 +\frac{1}{ivD + 2m_Q} i\fmslash{D}^{\bot} )h_{v}
\end{equation}  
Here $iD^{\bot} = iD - v (ivD)$ acts on the transverse degrees of freedom.
The expansion of (\ref{eq:lagMRRnl}) in powers of the inverse heavy quark
mass $m_Q$ yields the well known tree level lagrangian
\begin{equation}\label{eq:lagMRRl}
\mathcal L = \bar h_v  (ivD) h_v +  \frac{1}{2m_Q}\bar h_v i\fmslash{D}^{\bot}
\sum_{n=0}^{\infty}\biggl(\frac{-ivD}{2m_Q}\biggr)^n i\fmslash{D}^{\bot} h_v\,.
\end{equation}
However the procedure which leads to the lagrangian (\ref{eq:lagMRRl}) is not unique since one can switch to another
formulation of HQET by an appropriate field redefinition without
affecting matrix elements (see below). What formulation is chosen is more 
or less a matter of convenience. Therefore we will refer to the 
formulation of HQET based on the lagrangian (\ref{eq:lagMRRl}) as MRR--HQET.

In the presence of radiative corrections (\ref{eq:lagMRRl}) generalizes to
\begin{equation}\label{eq:lagMRRgen}
\mathcal L = \bar h_v (ivD) h_v + \sum_{n = 1}^{\infty}\frac{1}{(2m_{Q})^n}
\mathcal L^{(n)} \,,
\end{equation}
where  
\begin{equation}\label{eq:lagn}
\mathcal L^{(n)} = \sum_i C^{(n)}_i(\mu) \mathcal O^{(n)}_i(\mu)
\end{equation}
is a sum of operators multiplied by short distance coefficients.

A redefinition of the field $h_{v}$ implies a reordering of the 
$1/m_{Q}$--expansion and consequently a mixing of the 
coefficients. This means that in general the 
decomposition (\ref{eq:lagn}) is not unique but depends on what specific
formulation of HQET has been chosen. 
 
Under renormalization the set of operators in the tree level
lagrangian (\ref{eq:lagMRRl}) has to be completed to 
an operator basis. This basis in general has to include 
operators which vanish by the lowest order equation
of motion $(ivD)h_v = 0$ (EOM), since these operators
are needed as counterterms of physical operators.
At the highest order $\mathcal O(1/m_{Q})$ to which the $1/m_{Q}$--expansion
is extended one can forget about EOM--operators completely.
However, their presence at a certain lower $\mathcal O(1/m_{Q})$  affects
the coefficients in the orders beyond if the EOM--operators are removed 
from the lagrangian.
There exist several procedures how to remove these operators consistently,
each of them leading to the same unique minimal lagrangian
\footnote{The terminology ``minimal'' refers to the consistent usage
of $iD_\mu$ and $ivD$ to construct the operator basis. If instead 
$iD_\mu^\perp$ and $ivD$ are used EOM--operators are ``hidden''
in the $(ivD)$--piece of $iD_\mu^\perp$ and a decomposition
into operators vanishing or non--vanishing by the EOM is not
obvious.} 
which does not depend on the specific formulation of HQET one starts with.

In the present paper this will be done by an appropriate
field redefinition. However, first of all 
in every $\mathcal O(1/m_{Q})$ a complete off shell operator basis 
including EOM operators has to be renormalized.

In the present paper we restrict ourselves to the one loop renormalization 
of the operator bases appearing up to $\mathcal O(1/m_Q^3)$.
The set of operators up
to mass dimension 7 is given by 
\begin{equation}
\mathcal O^{(n)}_i = \bar h_v i\mathcal D^{(n)}_i h_v\,,
\end{equation}
where $i\mathcal D^{(n)}_i$ are the combinations of 
covariant derivatives defined in table 1.
To keep things simple we disregard four fermion operators and 
pure gluonic operators.
The lagrangian (\ref{eq:lagMRRgen}) can be written as
\newpage
\begin{eqnarray}\label{eq:MRRlag}
\mathcal L  &=& \bar h_{v} \bigg[ (ivD) 
         + \frac{1}{2m_{Q}} \sum_{i=1}^{3}C^{(1)}_{i}i\mathcal D^{(1)}_i
         + \frac{1}{(2m_{Q})^{2}} \sum_{i=1}^{7}C^{(2)}_{i}i\mathcal
         D^{(2)}_i
\nonumber \\
      &&  \hspace{4.5cm} + \frac{1}{(2m_{Q})^{3}} \sum_{i=1}^{24}C^{(3)}_{i}i\mathcal D^{(3)}_i \biggr] 
h_{v} \,.
\end{eqnarray}
\begin{table}
\begin{center}
\begin{tabular}{|c|c|c|c|}\hline
$i$&$i\mathcal D^{(n)}_i$ &tree level&coefficient of $(\alpha_{s}(\mu)/\pi)\ln(\mu/m_Q)$\\\hline
$i\mathcal D^{(1)}_1 $ & $(iD)^2 $ & $ 1$ & $0 $ \\
$i\mathcal D^{(1)}_2 $ & $-i\sigma_{\mu\nu}iD^\mu iD^\nu $ & $1 $ & $ 1/2C_{A} $ \\
$i\mathcal D^{(1)}_3 $ & $(ivD)^2 $ & $-1$ & $- 3 C_{F} -3/2 \xib C_{F} $ \\ \hline

$i\mathcal D^{(2)}_1 $ & $iD_\mu(ivD)iD^\mu $ & $ -1$ & $ 2/3C_{A} + 8/3C_{F} $ \\
$i\mathcal D^{(2)}_2 $ & $i\sigma_{\mu\nu}iD^\mu(ivD)iD^\nu $ & $1 $ & $ C_{A} $ \\
$i\mathcal D^{(2)}_3 $ & $(ivD)(iD)^2 $ & $ 0$ & $ -3/2\xib C_{F} - 1/3C_{A} - 13/3C_{F} $ \\
$i\mathcal D^{(2)}_4 $ & $(iD)^2(ivD) $ & $0 $ & $ -3/2\xib C_{F} - 1/3C_{A} - 13/3C_{F} $ \\
$i\mathcal D^{(2)}_5 $ & $(ivD)^3 $ & $1 $ & $9\xib C_{F} + 12C_{F} $ \\
$i\mathcal D^{(2)}_6 $ & $(ivD)i\sigma_{\mu\nu}iD^\mu iD^\nu $ & $0 $ & $ 3/2\xib C_{F} + 3C_{F} $ \\
$i\mathcal D^{(2)}_7 $ & $i\sigma_{\mu\nu}iD^\mu iD^\nu(ivD) $ & $0 $ & $ 3/2\xib C_{F} + 3C_{F} $ \\ \hline

$i\mathcal D^{(3)}_1 $ & $iD_{\mu}(ivD)^2 iD^{\mu} $ & $ 1$ & $ 3/2\xib C_{F} - 28/3C_{A} - 23/3C_{F}  $ \\
$i\mathcal D^{(3)}_2 $ & $ (iD)^2(iD)^2 $ & $ 0 $ & $ - 3/2\xib C_{F} - 7/6C_{A} - 17/3C_{F} $ \\
$i\mathcal D^{(3)}_3 $ & $iD_{\mu}(iD)^2 iD^{\mu} $ & $ 0 $ & $ - 3/2\xib C_{F} + 5C_{A} - 1/3C_{F} $ \\
$i\mathcal D^{(3)}_4 $ & $iD_{\mu}iD_{\nu} iD^{\mu}iD^{\nu} $ & $ 0 $ & $ 3/2\xib C_{F} - 23/6C_{A} + 3C_{F} $ \\
$i\mathcal D^{(3)}_5 $ & $i\sigma^{\mu\nu}iD_{\mu} (ivD)^2 iD_{\nu} $ & $-1 $ & $- 3/2\xib C_{F} + 8/3C_{A} - 1/3C_{F} $ \\
$i\mathcal D^{(3)}_6 $ & $i\sigma^{\mu\nu}iD_{\mu}iD_{\nu}(iD)^2 $ & $ 0 $ & $3/2\xib C_{F} - 7/6C_{A} + 13/3C_{F} $ \\
$i\mathcal D^{(3)}_7 $ & $i\sigma^{\mu\nu}iD_{\rho}iD_{\mu}iD_{\nu}iD^{\rho} $ & $ 0 $ & $ 3/2\xib C_{F} - 19/6C_{A} + 1/3C_{F} $ \\
$i\mathcal D^{(3)}_8 $ & $i\sigma^{\mu\nu}(iD)^2 iD_{\mu}iD_{\nu} $ & $ 0 $ & $3/2\xib C_{F} - 7/6C_{A} + 13/3C_{F} $ \\
$i\mathcal D^{(3)}_9 $ & $i\sigma^{\mu\nu}iD_{\mu}(iD)^2iD_{\nu} $ & $ 0 $ & $ 3/2\xib C_{F} - 5C_{A} + 3C_{F} $ \\
$i\mathcal D^{(3)}_{10} $ & $i\sigma^{\mu\nu}iD_{\mu}iD_{\rho}iD_{\nu}iD^{\rho} $ & $ 0 $ & $ - 3/2\xib C_{F} + 11/2C_{A} - 3C_{F} $ \\
$i\mathcal D^{(3)}_{11} $ & $i\sigma^{\mu\nu}iD_{\rho}iD_{\mu}iD^{\rho}iD_{\nu} $ & $ 0 $ & $ - 3/2\xib C_{F} + 11/2C_{A} - 3C_{F} $ \\
$i\mathcal D^{(3)}_{12} $ & $g_s^2 F^{a\mu\nu}F^a_{\mu\nu} $ & $ 0 $ & $1/12C_{A} $ \\
$i\mathcal D^{(3)}_{13} $ & $g_s^2 v_{\nu}v^{\rho}F^{a\mu\nu}F^a_{\mu\rho} $ & $ 0 $ & $ -1/3C_{A} $ \\
$i\mathcal D^{(3)}_{14} $ & $(ivD)^{4} $ & $ -1$ & $ - 89/2\xib C_{F} - 41C_{F} $ \\
$i\mathcal D^{(3)}_{15} $ & $(ivD)^{2}(iD)^{2} $ & $ 0 $ & $- 1/2\xib C_{A} + 15/2\xib C_{F} + 17/6C_{A} + 41/3C_{F} $ \\
$i\mathcal D^{(3)}_{16} $ & $ (ivD)(iD)^{2}(ivD) $ & $ 0 $ & $ -  \xib C_{A} + 19/2\xib C_{F} - 20/3C_{A} + 23/3C_{F} $ \\
$i\mathcal D^{(3)}_{17} $ & $(iD)^{2}(ivD)^{2} $ & $ 0 $ & $ - 1/2\xib C_{A} + 15/2\xib C_{F} + 17/6C_{A} + 41/3C_{F} $ \\
$i\mathcal D^{(3)}_{18} $ & $ iD_{\mu}(ivD)iD^{\mu}(ivD) $ & $ 0 $ & $\xib C_{A} + 31/6C_{A} + 10/3C_{F} $ \\
$i\mathcal D^{(3)}_{19} $ & $(ivD)iD_{\mu}(ivD)iD^{\mu} $ & $ 0 $ & $ \xib C_{A} + 31/6C_{A} + 10/3C_{F} $ \\
$i\mathcal D^{(3)}_{20} $ & $i\sigma^{\mu\nu}iD_{\mu}iD_{\nu}(ivD)^{2} $ & $ 0 $ & $ - 15/2\xib C_{F} - 5/6C_{A} - 37/3C_{F} $ \\
$i\mathcal D^{(3)}_{21} $ & $i\sigma^{\mu\nu}(ivD)iD_{\mu}iD_{\nu}(ivD) $ & $ 0 $ & $\xib C_{A} - 7/2\xib C_{F} + 41/6C_{A} - 11/3C_{F} $ \\
$i\mathcal D^{(3)}_{22} $ & $i\sigma^{\mu\nu}(ivD)^{2}iD_{\mu}iD_{\nu} $ & $ 0 $ & $ - 15/2\xib C_{F} - 5/6C_{A} - 37/3C_{F} $ \\
$i\mathcal D^{(3)}_{23} $ & $i\sigma^{\mu\nu}iD_{\mu}(ivD)iD_{\nu}(ivD) $ & $ 0 $ & $ - 37/6C_{A} - 14/3C_{F} $ \\
$i\mathcal D^{(3)}_{24} $ & $i\sigma^{\mu\nu}(ivD)iD_{\mu}(ivD)iD_{\nu} $ & $ 0 $ & $ - 37/6C_{A} - 14/3C_{F} $ \\ \hline
\end{tabular}
\end{center}
\caption{Operators up to $\mathcal O(1/m_Q^3)$ and their 
coefficients at tree level and $\mathcal O(\alpha_s)$.
The operators of mass dimensions 5, 6 and 7 are denoted $\mathcal O^{(1)}_i$,
$\mathcal O^{(2)}_i$ and $\mathcal O^{(3)}_i$ respectively
and the corresponding expressions have to be sandwiched 
between heavy quark spinors. We use $\bar \xi  = 1 - \xi $ as 
gauge parameter.}
\end{table}
The Wilson coefficients $C^{(n)}_{i}$ obey the renormalization group equation
\begin{equation}
\label{eq:RG}
\frac{d}{d\ln \mu}C^{(n)}_{i}(\mu) 
+(\frac{\alpha_{s}(\mu)}{\pi}) \sum_{j}\hat{\gamma}^{(3)\top}_{ij } C^{(3)}_{j}(\mu) = 0\,.
\end{equation}
Since we are working to leading logarithmic accuracy the one loop 
anomalous dimensions have been already plugged in.
The anomalous dimensions are well
known up to $\mathcal O(1/m_{Q}^{2})$ \cite{balzi,blok,bauer,finke} 
and  have recently 
been calculated 
at $\mathcal O(1/m_{Q}^{3})$ \cite{balzereit}.

The operator basis at 
$\mathcal O(1/m_{Q}^{3})$ is very large which makes an analytic solution of the
renormalization group equation (\ref{eq:RG}) 
too complicated. Therefore we extract 
from the exact solution 
\begin{equation}
\label{eq:exact}
C_{i}^{(n)}(\mu) =\sum_{j} \biggl[ \biggl( \frac{\alpha_s(\mu)}{\alpha_s(m_Q)}
                    \biggr)^{\frac{\hat{\gamma}^{(n)\top}}{2\beta^{(0)}}}
                    \biggr]_{ij}C^{(n)}_{j}(m_Q)
\end{equation} 
the first order 
logarithmic corrections in the Wilson coefficients:
 \begin{equation}\label{eq:logcorr}
C^{(n)}_{i}(\mu) = C^{(n)}_{i}(m_Q) 
           -\rho\,\,
             \sum_{j}\hat{\gamma}^{(3)\top}_{ij}C_{j}^{(n)}(m_Q)
           + \mathcal O((\frac{\alpha_{s}}{\pi})^2)
\end{equation}
Here $\rho = (\alpha_{s}(\mu)/\pi)\ln(\mu/m_Q)$ and 
$C^{(n)}_{i}(m_Q)$ is the tree level value of the coefficient at the 
matching scale.

In what follows the coefficients are subject of certain relations.
Since the first order logarithmic correction is unique
we assume that if the coefficients in the approximation (\ref{eq:logcorr})
obey the relations so will do the resummed coefficients. 

In order to remove the EOM--operators from the lagrangian (\ref{eq:MRRlag})
we now redefine the heavy quark field $h_{v}$ in terms of a ``physical''
field $h_{v}^{(p)}$
\begin{equation}\label{eq:redef}
h_v = \biggl[1 + \frac{1}{2m_Q}P^{(1)}(v,iD) + \frac{1}{(2m_Q)^2}P^{(2)}(v,iD)
          + \frac{1}{(2m_Q)^3}P^{(3)}(v,iD) \biggr]h_{v}^{(p)}\,,
\end{equation}
where
\begin{align}
P^{(1)}(v,iD) &= p^{(1)}_{1}(ivD) \\
P^{(2)}(v,iD) &= \sum_{i=1}^{3}p^{(2)}_{i} i\mathcal D^{(1)}_{i}\\
P^{(3)}(v,iD) &= \sum_{i=1}^{7}p^{(3)}_{i} i\mathcal D^{(2)}_{i}
\end{align}
with 
\begin{align}
p^{(1)}_{1} &=    1/2 +\rho( 3/4C_F\overline{\xi} + 3/2C_F) \nonumber\\
p^{(2)}_{1} &=  - 1/2 + \rho(3/4C_F\overline{\xi} + 17/6C_F + 1/3C_A) \nonumber\\
p^{(2)}_{2} &= - 1/2 +\rho( 3/4C_F\overline{\xi} + 3/2C_F - 1/4C_A)\nonumber \\
p^{(2)}_{3} &= - 1/8 - \rho(  27/8C_F\overline{\xi} + 15/4C_F)\nonumber \\ 
p^{(3)}_{1} &= 3/2  -\rho(  3/4C_F\overline{\xi} + 37/6C_F + C_A\overline{\xi} + 9/2C_A)\nonumber \\
p^{(3)}_{2} &= - 3/2 +\rho( 3/4C_F\overline{\xi} + 37/6C_F + 14/3C_A)\nonumber \\
p^{(3)}_{3} &= - 1/8 + \rho( - 27/8C_F\overline{\xi} - 85/12C_F + 1/2C_A\overline{\xi} - 5/2C_A)\nonumber\\
p^{(3)}_{4} &=  - 5/8 + \rho( - 29/8C_F\overline{\xi} - 11/12C_F + 1/2C_A\overline{\xi} + 3C_A)\nonumber\\
p^{(3)}_{5} &=  1/16 +\rho( 505/32C_F\overline{\xi} + 193/16C_F)\nonumber \\
p^{(3)}_{6} &=  1/8 +\rho( 27/8C_F\overline{\xi} + 85/12C_F + 43/48C_A)\nonumber \\
p^{(3)}_{7} &=  5/8 +\rho( 5/8C_F\overline{\xi} - 5/12C_F - 1/2C_A\overline{\xi} - 137/48C_A )\label{eq:pin} \,.
\end{align}
In what follows we also need the inverse to (\ref{eq:redef})
\begin{equation}\label{eq:redefinv}
h^{(p)}_v = \biggl[1 + \frac{1}{2m_Q}Q^{(1)}(v,iD) + \frac{1}{(2m_Q)^2}Q^{(2)}(v,iD)
          + \frac{1}{(2m_Q)^3}Q^{(3)}(v,iD) \biggr]h_{v} 
\end{equation}
with
\begin{align}\label{eq:redefinvc}
Q^{(1)}(v,iD) &= -P^{(1)}(v,iD) \nonumber\\
Q^{(2)}(v,iD) &= -P^{(2)}(v,iD) + (P^{(1)}(v,iD))^{2} \\
Q^{(3)}(v,iD) &= -P^{(3)}(v,iD) + \{P^{(1)}(v,iD),P^{(2)}(v,iD)\} 
- (P^{(1)}(v,iD))^{3} \,.
\nonumber 
\end{align}
In (\ref{eq:pin}) the coefficients $p^{(n)}_{i}$ are choosen such 
that under application of (\ref{eq:redef}) up to $\mathcal O(1/m_{Q}^{3})$
all operators vanishing by the EOM are removed from the lagrangian
(\ref{eq:MRRlag}) and we are left with
\begin{eqnarray}\label{eq:physlag}
\mathcal L^{(p)} &=& \bar h_{v}^{(p)}\bigg[ (ivD) 
         + \frac{1}{2m_{Q}} \sum_{i=1}^{2}C^{(1)p}_{i}i\mathcal D^{(1)}_i
         + \frac{1}{(2m_{Q})^{2}} \sum_{i=1}^{2}C^{(2)p}_{i}i\mathcal
         D^{(2)}_i
\nonumber \\
      &&  \hspace{4.5cm} + \frac{1}{(2m_{Q})^{3}} \sum_{i=1}^{13}C^{(3)p}_{i}i\mathcal D^{(3)}_i \biggr] 
h_{v}^{(p)}
\end{eqnarray}
with the transformed coefficients listed in table 2.
\begin{table}[t]
\begin{center}
\begin{tabular}{|c|c|c|} \hline
$C^{(n)p}_i$& tree level & 
coefficient of $\rho$\\ \hline
$C^{(1)p}_1$ &$1$  & $0$ \\
$C^{(1)p}_2$ &$1$  & $C_{A}/2$ \\ \hline
$C^{(2)p}_1$ &$-1$  & $2/3C_{A}+8/3C_{F}$ \\
$C^{(2)p}_2$ &$1$  & $C_{A}$ \\ \hline
$C^{(3)p}_1$ & $2$ & $-25/3\,C_A -32/3\,C_F$ \\ 
$C^{(3)p}_2$ & $-1$ & $-1/2\,C_A $ \\ 
$C^{(3)p}_3$ & $-1$  & $4\,C_A +8/3\,C_F$ \\ 
$C^{(3)p}_4$  & $1$ & $-17/6\,C_A $ \\ 
$C^{(3)p}_5$  & $-2$ & $5/3\,C_A+ 8/3\,C_F$ \\ 
$C^{(3)p}_6$ & $1$ & $-\,C_A$ \\ 
$C^{(3)p}_7$ & $1$  & $-13/6\,C_A -8/3\,C_F$ \\ 
$C^{(3)p}_8$ & $1$ & $-\,C_A $ \\ 
$C^{(3)p}_9$ & $1$ & $-4\,C_A $ \\ 
$C^{(3)p}_{10}$ & $-1$  & $9/2\,C_A $\\ 
$C^{(3)p}_{11}$ & $-1$  & $9/2\,C_A $ \\ 
$C^{(3)p}_{12}$ & $0$  & $1/12\,C_A$  \\ 
$C^{(3)p}_{13}$& $0$ & $-1/3\,C_A$ \\ \hline
\end{tabular}
\caption{Coefficients of the physical operator basis.}
\end{center}
\end{table}
Comparison with table 1  shows 
that the coefficients $C^{(1/2)p}_{1/2}$ of the physical operators
at $\mathcal O(1/m_{Q})$ and $\mathcal O(1/m_{Q}^{2})$  are
not affected by the redefinition (\ref{eq:redef}) whereas 
the coefficients at $\mathcal O(1/m_{Q}^{3})$  
are modified. In particular new operators appear already 
at tree level and the one loop contributions  are now independent of the 
gauge parameter, which is a crucial property of  
coefficients of physical operators. Note that in order to remove the 
EOM operators appearing at one loop order in the renormalized lagrangian
one has to include terms of $\mathcal O(\alpha_{s})$ in the field redefinition
(\ref{eq:redef}). In other words the tree level field redefinition 
will not suffice to remove all EOM operators from the lagrangian
once radiative corrections are included. 

Let us shortly comment on other methods which may reduce the 
MRR--lagrangian (\ref{eq:MRRlag})
to the minimal lagrangian (\ref{eq:physlag}).

On the level of matrix elements the reduction effectively takes place 
if insertions
of the EOM--operators in the lagrangian (\ref{eq:MRRlag}) into time ordered products are properly
taken care of by contraction identities \cite{balzereit} 
\begin{equation}
i\Tprod{\bar h_v F(iD) (ivD)h_v, \bar h_v G(iD) h_v} = 
  - \bar h_v F(iD) G(iD) h_v + \ldots
\end{equation}
which allow to remove these unphysical T--products  in favour of 
local operators. This procedure corresponds to a reorganisation of 
the $1/m_{Q}$--expansion and is equivalent to a field redefinition.

An alternative method applies the full EOM
of the heavy quark, i.e. the EOM derived from the lagrangian
(\ref{eq:MRRlag})
including power corrections
\begin{equation}
(ivD) h_v = \mathcal O(1/m_Q)\,,
\end{equation}
to the EOM operators in the MRR--lagrangian itself.
This way these operators are related to operators of
higher dimension and the $1/m_Q$--expansion and the 
coefficients 
are reshuffeld. 
We have checked explicitly that all three methods,
i.e. field redefinition, the use of contraction identities and application of
the full EOM,
yield the same minimal lagrangian (\ref{eq:physlag}).
\section{Reparametrization Invariance and Field Redefinitions}
\label{sec:RI}
The Greensfunctions of QCD only depend on the full heavy quark 
momentum independent from its
decomposition into on--shell component $m_{Q}v$ and  
residual heavy quark momentum in terms 
of which HQET is defined. This is reflected in HQET 
by a new symmetry, the reparametrization invariance.
A change of the velocity $v \to v + \delta v$, $v\cdot \delta v =0$,
is accompanied by an appropriate change of the 
heavy quark field $h_{v}\to h_{v+\delta v} = h_{v} + \delta h_{v}$
such that the HQET lagrangian is invariant. 
In general the field transformation $\delta h_{v} $ depends on the specific 
formulation of HQET. In this section we derive a description 
how the transformation has to be modified if we switch to a different
formulation of HQET by means of a field redefinition.  

The MRR--lagrangian (\ref{eq:MRRlag}) is unphysical in the sense,
that it contains operators vanishing by the heavy quark EOM
which have to be removed carefully to extract the final minimal 
lagrangian (\ref{eq:physlag}). What makes the MRR--lagrangian interesting
is that it deduces from the QCD lagrangian in the
most direct and simple way. 
In order to investigate reparametrization invariance
which is a genuine property of QCD itself we therefore 
study first of all the MRR--lagrangian. 

Suppose the lagrangian in the MRR formulation 
$\mathcal L(v,h_{v},\bar h_{v})$ 
is invariant under the reparametrization transformation 
\begin{equation}\label{eq:RItrafoMRR}
v \to v + \delta v \qquad  h_{v} \to h_{v} + \delta h_{v} \qquad 
\delta h_{v} = F(v,\delta v,iD)h_{v} \,.
\end{equation}
In general $F(v,\delta v,iD)$ has an expansion in powers
$1/m_{Q}$ and $\alpha_{s}$. Its concrete form is irrelevant
for the more general consideration in this section and will 
be specified in section \ref{sec:RIMRR}.

In terms of the lagrangian reparametrization invariance means
\begin{equation}\label{eq:RIm}
\biggl[
\delta v \frac{\partial}{\partial v} 
+\frac{\delta}{\delta h_{v}} \delta h_{v}
+ \delta \bar h_{v} \frac{\delta}{\delta \bar h_{v}} 
\biggr] \mathcal L(v,h_{v},\bar h_{v}) = 0\,.
\end{equation} 

Given (\ref{eq:RItrafoMRR}) we want to derive an analogous 
transformation which leaves the minimal lagrangian
$\mathcal L^{(p)}(v,h^{(p)}_{v},\bar h^{(p)}_{v})$ (\ref{eq:physlag})
invariant:
\begin{align}
v &\to  v + \delta v \\
h^{(p)}_{v} &\to h^{(p)}_{v} + \delta h^{(p)}_{v} 
\end{align}
The field redefinition
\begin{equation}\label{eq:redefgen}
h_{v} = [ 1 + P(v,iD)]h^{(p)}_{v} \leftrightarrow 
h^{(p)}_{v} = [ 1 + Q(v,iD)]h_{v}\,,
\end{equation}
where $P(v,iD)$, $Q(v,iD)$ have to be identified with the
terms in square brackets in (\ref{eq:redef},\ref{eq:redefinv})
connects both formulations of HQET:
\begin{equation}\label{eq:connlag}
\mathcal L(v,h_{v},\bar h_{v}) = 
\mathcal L^{(p)}\left(v, [ 1 + Q(v,iD)]h_{v},\bar h_{v} [ 1 + \bar Q(v,iD)]
\right)
\end{equation}
Inserting (\ref{eq:connlag}) in (\ref{eq:RIm}) and using 
(\ref{eq:redefgen}) we derive
\begin{equation}
\biggl[
\delta v \frac{\partial}{\partial v} 
+\frac{\delta}{\delta h^{(p)}_{v}} \delta h^{(p)}_{v}
+ \delta \bar h^{(p)}_{v} \frac{\delta}{\delta \bar h^{(p)}_{v}} 
\biggr] \mathcal L^{(p)}(v,h^{(p)}_{v},\bar h^{(p)}_{v}) = 0
\end{equation}
where now
\begin{equation}\label{eq:RItrafophys}
\delta h^{(p)}_{v} = 
\biggl(
[1 + Q(v,iD)] F(v,\delta v,iD) 
+ [\frac{\partial}{\partial v}Q(v,iD)] 
\biggr)[1 + P(v,iD)] h^{(p)}_{v} \,.
\end{equation}
Inserting the expressions for $F(v,\delta v,iD)$,
$P(v,iD)$ and $Q(v,iD)$ and expanding to the 
required order $\mathcal O(1/m_{Q})$ and $\mathcal O(\alpha_{s})$
yields the reparametrization 
transformation appropriate to  the minimal lagrangian (\ref{eq:physlag}).
In the next sections we derive this transformation at tree--level
and $\mathcal O(\alpha_{s})$ and show explicitly that 
the minimal lagrangian (\ref{eq:physlag}) is reparametrization invariant.

\section{Reparametrization Invariance of the MRR--Lagrangian}
\label{sec:RIMRR}
From the reparametrization invariance of the 
tree level matching condition (\ref{eq:matchnl}) 
\begin{equation}
\delta v \frac{d}{dv}Q(x)  = 0
\end{equation}
we derive the reparametrization transformation  at tree level:
\begin{equation}\label{eq:RItrafotree}
\delta h_{v} = \biggl[
im_{Q}\delta vx +  \frac{\delta \fmslash{v}}{2}
+ \frac{\delta \fmslash{v}}{2}\frac{i\fmslash{D}^{\bot}}{2m_{Q}}
- \frac{\delta \fmslash{v}}{2}\frac{(ivD)i\fmslash{D}^{\bot}}{(2m_{Q})^{2}}
\biggr]h_{v}
\end{equation}
The second term in the square brackets is irrelevant since
$P^{+}_{v}\delta \fmslash{v}P^{+}_{v} = 0$.
We have included in the MRR--lagrangian (\ref{eq:MRRlag}) terms
up to $\mathcal O(1/m_{Q}^{3})$. Since $\delta h_{v} = \mathcal O(m_{Q})$
the field reparametrization (\ref{eq:RItrafotree})  must be known up to
$\mathcal O(1/m_{Q}^{2})$.
A short calculation shows that the tree level MRR--lagrangian, 
i.e. (\ref{eq:MRRlag}) with $\rho =0$ is invariant under 
the transformation (\ref{eq:RItrafotree}),
i.e. its variation is $\mathcal O(1/m_{Q}^{3})$.
The question is now, if (\ref{eq:RItrafotree}) is 
the correct transformation formula if radiative corrections 
are included in the lagrangian. 
In fact the tree level transformation (\ref{eq:RItrafotree}) leaves the 
one loop effective lagrangian invariant as long as 
power corrections not higher than $\mathcal O(1/m_{Q}^{2})$ 
are included. In this case the term of $\mathcal O(1/m_{Q}^{2})$
in (\ref{eq:RItrafotree}) is irrelevant. 
However this situation changes drastically if power corrections
of $\mathcal O(1/m_Q^3)$ or higher are taken into
account.
The inconsistency shows up if we require reparametrization invariance 
of the lagrangian
(\ref{eq:MRRlag}) under the tree level transformation 
(\ref{eq:RItrafotree}) to all orders in the strong coupling. Then
the shortdistance coefficients
$C^{(n)}_{i}$, $n = 1, 2, 3$, must fulfill 
the following relations:
\begin{align}
C^{(1)} &= 1+\Delta_{1} \nonumber \\
C^{(2)}_{2} & =2C^{(1)}_{2} - 1+\Delta_{2} \nonumber \\
2C^{(2)}_{4} + C^{(2)}_{1} & =  2C^{(1)}_{3} + 1+\Delta_{3}\nonumber \\
2C^{(2)}_{5} + C^{(2)}_{1} & =  2C^{(1)}_{3} + 1+\Delta_{4}\nonumber \\
C^{(3)}_{5} + C^{(3)}_{24} &= -2C^{(2)}_{6}-2C^{(2)}_{2} -C^{(1)}_{2}
-C^{(1)}_{3} + 1 +\Delta_{5}\nonumber \\
C^{(3)}_{23}-C^{(3)}_{24} &= 2C^{(2)}_{6}-2C^{(2)}_{7}+\Delta_{6} \nonumber \\
C^{(3)}_{5} + C^{(3)}_{23} &= -2C^{(2)}_{7}-2C^{(2)}_{2} -C^{(1)}_{2}
-C^{(1)}_{3} + 1 +\Delta_{7}\nonumber \\
C^{(3)}_{11}+2C^{(3)}_{8}+C^{(3)}_{7} &= 2C^{(2)}_{6}+\Delta_{8} \nonumber \\
C^{(3)}_{11}+C^{(3)}_{10}+2C^{(3)}_{9} &= 2C^{(2)}_{2} -2C^{(1)}_{2}+\Delta_{9} \nonumber \\
C^{(3)}_{10}+C^{(3)}_{7}+2C^{(3)}_{6} &= 2C^{(2)}_{7} +\Delta_{10}
\label{eq:MRRrel}\\
C^{(3)}_{18}+2C^{(3)}_{17}+C^{(3)}_{1} &= 2C^{(2)}_{5} +
C^{(1)}_{3}+C^{(1)}_{2}-1 +\Delta_{11} \nonumber \\
C^{(3)}_{19}+C^{(3)}_{18}+2C^{(3)}_{16} &= 2C^{(2)}_{5} -2C^{(1)}_{2} 
+\Delta_{12}\nonumber \\
C^{(3)}_{19}+2C^{(3)}_{15}+C^{(3)}_{1} &= 2C^{(2)}_{5}
+C^{(1)}_{3}+C^{(1)}_{2} -1 +\Delta_{13}\nonumber \\
C^{(3)}_{4}+C^{(3)}_{3}+2C^{(3)}_{2} &= 2C^{(2)}_{3} +C^{(1)}_{1}-C^{(1)}_{2}+\Delta_{14} \nonumber \\
C^{(3)}_{4}+C^{(3)}_{3} &= C^{(2)}_{1}+ C^{(1)}_{2}+\Delta_{15} \nonumber \\
C^{(3)}_{4}+C^{(3)}_{3} +2C^{(3)}_{2}  &= 2C^{(2)}_{4}
+C^{(1)}_{1}-C^{(1)}_{2} +\Delta_{16} \nonumber \\
C^{(3)}_{10}+C^{(3)}_{9}  &= C^{(1)}_{2}-C^{(1)}_{1}+\Delta_{17} \nonumber \\
C^{(3)}_{11}-C^{(3)}_{10} &= \Delta_{18} \nonumber \\
C^{(3)}_{11}+C^{(3)}_{9} &= C^{(1)}_{2} -C^{(1)}_{1}+\Delta_{19}  \nonumber 
\end{align}
If the tree level reparametrization transformation (\ref{eq:RItrafotree}) 
is the correct transformation even in the presents of radiative corrections 
all $\Delta_{i}$ have to
vanish. A deviation form this value would signal 
reparametrization invariance breaking.
Insertion of the coefficients in table 1 
shows that 5 of the relations involving coefficients of operators 
at $\mathcal O(1/m_{Q}^{3})$ are violated at $\mathcal O(\alpha_{s})$:
\begin{align}
\Delta_{5} = \Delta_{7}= -\Delta_{11}= -\Delta_{13} &= - \rho (C_{A} + 2 C_{F})\nonumber
\\
\Delta_{12} &= -\rho(2C_{A} + 2C_{F} - \xib C_{F})\\
\Delta_{i} &= 0 \quad i \neq 5, 7, 11, 12, 13\nonumber
\end{align}
Obviously the naive understanding of reparametrization invariance
that the tree level transformation remains correct even 
if radiative corrections are included in the lagrangian
fails beyond $\mathcal O(1/m_Q^2)$.
However, the violation of reparametrization invariance 
can be cured if the reparametrization transformation itself is renormalized,
i.e. is supplemented by $\mathcal O(\alpha_s)$--corrections. 
Since the tree level transformation works well up to $\mathcal O(1/m_{Q}^{2})$ 
the term of $\mathcal O(1/m_{Q}^{2})$ in (\ref{eq:RItrafotree})
which is responsible for the relations between $\mathcal O(1/m_{Q}^{3})$
and the lower orders 
has to be corrected at $\mathcal O(\alpha_{s})$.
It comes out that  (\ref{eq:RItrafotree}) generalizes to
\begin{equation}\label{eq:RItrafo1}
\delta h_{v}  = \biggl[ 
im_{Q}\delta vx
+ \frac{1}{2m_{Q}} F_{1}
+ \frac{1}{(2m_{Q})^{2}}F_{2} 
\biggr]h_{v}
\end{equation}
where
\begin{align}\label{eq:FsMRR}
F_{1} &=c_{1}\frac{\delta \fmslash{v}}{2}i\fmslash{D} \\
F_{2} &=c_{21} \frac{\delta \fmslash{v}}{2}(ivD)i\fmslash{D}
+ c_{22} \frac{\delta \fmslash{v}}{2}i\fmslash{D}(ivD)
+  c_{23}(i\delta v D)(ivD)
+  c_{24}(ivD)(i\delta v D)
\end{align}
with
\begin{align}
c_{1} &= 1\nonumber\\
c_{21} &=-1  +\rho( C_{A}+ 2C_{F})\nonumber\\
c_{22} &=  \rho ( - C_{A} +\frac{1}{2}\overline{\xi}C_{F} - C_{F})\nonumber\\
c_{23} &= c_{24} = 0 \nonumber \,.
\end{align}
In (\ref{eq:RItrafo1}) the expression in square brackets coincides with
the operator $F(v,\delta v,iD)$ in (\ref{eq:RItrafoMRR}).
One may argue that one can always find a transformation formular similar to 
(\ref{eq:RItrafo1}) which leaves the lagrangian invariant whatever values
the coefficients $C_{i}^{(n)}$ take.
However this argument fails, since in the 
transformation formula at $\mathcal O(1/m_{Q}^{2})$ 
there are in general only 4 terms which can be 
modified to correct all possible violated relations in (\ref{eq:MRRrel}).  
For 
example in our case  
the correction of only two terms in the transformation formula 
suffices to recover reparametrization invariance.
\section{Switching to the Minimal Operator Basis}
\label{sec:RIphys}
Now we are in the position to combine the results of the previous sections
to calculate the reparametrization transformation of the 
``physical'' heavy quark field $h_{v}^{(p)}$.
Inserting (\ref{eq:redef},\ref{eq:redefinv}) and (\ref{eq:RItrafo1})
into (\ref{eq:RItrafophys}) and keeping only terms up to
$\mathcal O(1/m_{Q}^{2})$ and $\mathcal O(\alpha_{s})$ we get
\begin{equation}\label{eq:RItrafophys1}
\delta h_{v}^{(p)} = \biggl[ 
im_{Q}\delta vx
+ \frac{1}{2m_{Q}} F^{(p)}_{1}
+ \frac{1}{(2m_{Q})^{2}}F^{(p)}_{2} 
\biggr]h_{v}^{(p)}
\end{equation}
where
\begin{align}\label{eq:Fsphys}
F^{(p)}_{1} &=c^{(p)}_{11}\frac{\delta \fmslash{v}}{2}i\fmslash{D}
+c^{(p)}_{12}i\delta vD  \\
F^{(p)}_{2} &=c^{(p)}_{21} \frac{\delta \fmslash{v}}{2}(ivD)i\fmslash{D}
+ c^{(p)}_{22} \frac{\delta \fmslash{v}}{2}i\fmslash{D}(ivD)
+  c^{(p)}_{23}(i\delta v D)(ivD)
+  c^{(p)}_{24}(ivD)(i\delta v D)
\end{align}
with
\begin{align}
c^{(p)}_{11} &= 1\nonumber \\
c^{(p)}_{12} &=-1 +\frac{1}{3} \rho(C_{A}+4C_{F})\nonumber \\
c^{(p)}_{21} &=-1-\rho( \frac{25}{6}C_{A}+ \frac{8}{3}C_{F})
\nonumber \\
c^{(p)}_{22} &=  \rho(\frac{1}{2}\overline{\xi}C_{F}+ \frac{25}{16} C_{A}
+ \frac{11}{3}C_{F}) \label{eq:cphys}\\
c^{(p)}_{23} &=\frac{1}{2}-\rho(\frac{1}{4}\overline{\xi}C_{F}
+ \frac{11}{6}C_{A}+ \frac{11}{6}C_{F})\nonumber\\
c^{(p)}_{24} &= 1-\rho(\frac{7}{3}C_{A}+ 4 C_{F})\nonumber \,.
\end{align}
Note that the gauge parameter dependent terms in (\ref{eq:cphys})
cancels if  the transformation formula is applied to the 
lagrangian.
We have checked that the minimal lagrangian (\ref{eq:physlag}) 
is indeed invariant under (\ref{eq:RItrafophys1}).

The reparametrization transformation (\ref{eq:RItrafophys1}) 
of the physical heavy quark field 
receives one loop corrections already
at $\mathcal O(1/m_{Q})$ which is not the case 
for the transformation (\ref{eq:RItrafo1})
of the MRR--lagrangian. From this 
we expect, that to one loop order reparametrization invariance 
of the minimal lagrangian is already broken at $\mathcal O(1/m_{Q}^{2})$.
Let us therefore 
in analogy to section \ref{sec:RIMRR} derive the  relations among 
the coefficients $C^{(n)p}_{i}$ which
have to hold if we require the minimal lagrangian  (\ref{eq:physlag}) 
to be invariant under the tree level reparametrization transformation
(i.e. (\ref{eq:RItrafophys1}) with $\rho =0$ ) to
all orders in perturbation theory:
\begin{align}
C^{(1)p}_{1} &= 1+ \Delta^{(p)}_{1}\nonumber\\
C^{(2)p}_{2} &= 2C^{(1)p}_{2} - 1+ \Delta^{(p)}_{2} \nonumber\\
C^{(2)p}_{1} &= -1+\Delta^{(p)}_{3} \nonumber\\
C^{(3)p}_{5} &= -2C^{(2)p}_{2} -C^{(1)p}_{2} +1+\Delta^{(p)}_{4} \nonumber\\
C^{(3)p}_{11}+ 2C^{(3)p}_{8}+C^{(3)p}_{7} &= 2C^{(1)p}_{2}+\Delta^{(p)}_{5} \nonumber\\
C^{(3)p}_{11}+C^{(3)p}_{10}+2C^{(3)p}_{9} &= 2C^{(2)p}_{2} -2C^{(1)p}_{2}+\Delta^{(p)}_{6} \nonumber\\
C^{(3)p}_{10}+C^{(3)p}_{7}+2C^{(3)p}_{6} &= 2C^{(1)p}_{2}+\Delta^{(p)}_{7} \nonumber\\
C^{(3)p}_{1} &= C^{(1)p}_{2} + 1+\Delta^{(p)}_{8} \label{eq:physrel}\\
0&= C^{(1)p}_{2} - 1+\Delta^{(p)}_{9} \nonumber\\
C^{(3)p}_{4}+C^{(3)p}_{3}+2C^{(3)p}_{2} &= -C^{(1)p}_{1}-C^{(1)p}_{2}+\Delta^{(p)}_{10} \nonumber\\
C^{(3)p}_{4}+C^{(3)p}_{3} &= C^{(2)p}_{1}+ C^{(1)p}_{2}+\Delta^{(p)}_{11} \nonumber\\
C^{(3)p}_{10}+C^{(3)p}_{9} &= C^{(1)p}_{2}-C^{(1)p}_{1}+\Delta^{(p)}_{12} \nonumber\\
C^{(3)p}_{11}-C^{(3)p}_{10} &= \Delta^{(p)}_{13} \nonumber\\
C^{(3)p}_{11}+C^{(3)p}_{9}  &= C^{(1)p}_{2}-C^{(1)p}_{1}+\Delta^{(p)}_{14} \nonumber
\end{align} 
Invariance under the tree level reparametrization transformation 
requires $\Delta_{i}^{(p)} =0$. 
Inserting the coefficents $C^{(n)p}_{i}$ from table 2
shows that all relations hold at tree level
but some are violated at $\mathcal O(\alpha_{s})$:
\begin{align}
\Delta_{4} &= \rho(25/6 C_{A} + 8/3C_{F}) \nonumber\\
\Delta_{3} = -\Delta_{5} = -\Delta_{7} = \Delta_{10} &= \rho(2/3C_{A} +
8/3C_{F})\nonumber \\
\Delta_{8} &= -\rho(53/6C_{A} + 32/3 C_{F}) \\
\Delta_{9} &= -\rho C_{A}/2\nonumber\\
\Delta_{i} &=0 \quad i\neq 3, 4, 5, 7, 8, 9, 10\nonumber
\end{align}
Again the naive understanding
of reparametrization invariance fails and in order to maintain 
reparametrization invariance one has to add 
corrections of $\mathcal O(\alpha_{s})$ to the tree level
transformation. We have explicitly checked that
(\ref{eq:RItrafophys1}) leaves the minimal lagrangian 
invariant if the one loop corrections in (\ref{eq:cphys}) 
are included in the transformation.

In case of the MRR--lagrangian 
the naive application of reparametrization invariance 
accidentally works well up to $\mathcal O(1/m_{Q}^{2})$, 
while in case of the minimal lagrangian
it leads
to inconsistencies already appearing at lower orders.
Namely the third and  the ninth relation in (\ref{eq:physrel}) would imply
non renormalization of the operator $\mathcal O^{(2)}_{1}$ 
(Darwin operator) and the chromomagnetic operator $\mathcal O^{(1)}_{2}$.
However, the usual renormalization group approach yields
non vanishing radiative corrections for the coefficients of these
operators.

\section{Conclusions}
\label{sec:conc}
We have extended the analysis of the reparametrization invariance
of HQET to $\mathcal O(1/m_Q^3)$. We have shown explicitly
that the reparametrization transformation depends
on the specific formulation of HQET and that 
the transformation has to be modified if one switches 
to another formulation. As examples we have studied 
the MRR--formulation and the minimal lagrangian.

The main subject of this paper, however, was
the question if and in what sense reparametrization invariance
is maintained under renormalization.
We have demonstrated explicitly that the naive understanding of
reparametrization invariance -- application of the tree level
transformation to the renormalized lagrangian --
fails if higher orders $1/m_Q$  as well as radiative corrections are included.
In the case of the MRR--lagrangian accidentally the tree--level
transformation works well up to $\mathcal O(1/m_Q^2)$ but
fails at $\mathcal O(1/m_Q^3)$. By  contrast, in case of the
physical lagrangian an inconsistency already appears
at $\mathcal O(1/m_Q^2)$.

In order to 
recover reparametrization invariance of both
the MRR--lagrangian (\ref{eq:MRRlag}) and the
physical lagrangian (\ref{eq:physlag}) in the presence of
radiative corrections the corresponding
field transformations $\delta h_{v}$ and $\delta h_{v}^{(p)}$ 
have to be renormalized properly. However, given the correct
transformation formula in one formulation of HQET
the corresponding transformation in any other
formulation related to the first by means of a field redefiniton 
is fixed by (\ref{eq:RItrafophys}). 

To summarize, the application of reparametrization invariance at higher 
orders of the $1/m_Q$--expansion in order to extract short 
distance coefficients from lower order calculations has to be 
treated with care, since the transformation prescription depends on what
specific formulation of HQET has been chosen and 
requires renormalization if radiative corrections are included in the
effective lagrangian.

\section*{Acknowledgements}
This work was supported by   
the ``Forschergruppe: Quantenfeldtheorie, Computeralgebra und 
Monte Carlo Simulationen'' of the Deutsche Forschungsgemeinschaft.   



\begin{thebibliography}{10}
\bibitem{HQET_classic}
  N. Isgur and M. Wise, Phys.~Lett. {\bf B208},  504  (1988);
  N. Isgur and M. Wise, Phys.~Lett. {\bf B232},  113  (1989);
  E. Eichten and B. Hill, Phys.~Lett. {\bf B234},  511  (1990);
  H. Georgi, Phys.~Lett. {\bf B240},  447  (1990);
  B. Grinstein, Nucl.~Phys. {\bf B339},  253  (1990).
\bibitem{HQET_review}
  B. Grinstein, Ann. Rev. Nucl. Part. Sci. {\bf 42},  101  (1993);
  T. Mannel,  in {\em {QCD} -- 20 years later, Proceedings of the Workshop,
    Aachen 1992}, edited by P. Zerwas and H. Kastrup (World Scientific,
    Singapore, 1993);
  M. Neubert, Phys.~Rep. {\bf 245},  259  (1994).
\bibitem{repar}
  M. Luke and A. Manohar, Phys.~Lett. {\bf B286},  348  (1992).
\bibitem{blok}
  B. Blok, J.~G. K{\"o}rner, and D. Pirjol and J.C. Rojas, 
  Nucl.~Phys. {\bf B496}, 358 (1997).
\bibitem{grozin}
  A.~Czarnecki and A.G. Grozin, Phys.~Lett. {\bf B405}, 142 (1997).
\bibitem{manohar}
  A.V. Manohar, Phys.~Rev. {\bf D56}, 230 (1997).
\bibitem{finke}
  M.~Finkemeier and M.~McIrvin, Phys.~Rev. {\bf D55}, 377 (1997).
\bibitem{balzereit} 
  C. Balzereit, hep-ph 9801436.
\bibitem{MRR}
  T. Mannel, W. Roberts and Z. Ryzak, Nucl. Phys. {\bf B368}, 204 (1992).
\bibitem{balzi}
  C.~Balzereit and T.~Ohl, Phys.~Lett. {\bf B386}, 335 (1996).
\bibitem{bauer}
  C.~Bauer and A.V. Manohar, Phys.~Rev. {\bf D57}, 337 (1998). 
\end{thebibliography}
\end{document}